\documentclass[aps,prb,twocolumn,floatfix]{revtex4-2}
\usepackage[dvipdfmx]{graphicx} 
\usepackage[dvipdfmx]{color}
\usepackage{array}
\usepackage{mathrsfs}
\usepackage{ulem} 
\usepackage{amsmath} 
\usepackage{braket}
\usepackage{amsmath}

\begin{document}
\title{Six magnetization plateau phases in a spin-1/2 distorted kagome antiferromagnet: application to $\rm{Y}_3\rm{Cu}_9(\rm{OH})_{19}\rm{Cl}_8$}

\author{Kazu Bodaiji}
\affiliation{Department of Physics and Astronomy, Faculty of Science and Technology, Tokyo University of Science, Chiba 278-8510, Japan}
\author{Katsuhiro Morita}
\email[e-mail:]{katsuhiro.morita@rs.tus.ac.jp}
\affiliation{Department of Physics and Astronomy, Faculty of Science and Technology, Tokyo University of Science, Chiba 278-8510, Japan}
\author{Yoshiyuki Fukumoto}
\affiliation{Department of Physics and Astronomy, Faculty of Science and Technology, Tokyo University of Science, Chiba 278-8510, Japan}

\begin{abstract}
 A recently discovered kagome antiferromagnet $\rm{Y}_3\rm{Cu}_9(\rm{OH})_{19}\rm{Cl}_8$ has attracted significant interest due to its unique kagome lattice structure and magnetic properties.
The kagome lattice has three types of exchange interactions: one hexagonal coupling and two different triangular couplings. Previous studies have shown that its ground state is significantly different from that predicted for the undistorted kagome lattice, forming a coplanar spin state with a commensurate magnetic wave vector ${\mathbf Q}=(1/3,1/3)$.
  Two separate studies have proposed distinct sets of exchange interaction parameters for this compound.
  In this study, we investigate the ground state of the spin-1/2 Heisenberg kagome model with three types of nearest-neighbor exchange interactions under a magnetic field by exact diagonalization using the Lanczos method.  We reveal that clear magnetization plateaus at  $M/M_{\rm sat}$=1/3, 5/9, and 7/9 are present under both parameter sets, which are identified as magnon crystal states based on their spin structures.
   Our findings suggest that these plateaus could potentially be experimentally confirmed with magnetization measurements on $\rm{Y}_3\rm{Cu}_9(\rm{OH})_{19}\rm{Cl}_8$ under a magnetic field of approximately 300 T, achievable with state-of-the-art magnetic field generators. 
In order to get a deeper understanding of magnetism of $\rm{Y}_3\rm{Cu}_9(\rm{OH})_{19}\rm{Cl}_8$, we perform additional calculations by varying these interactions. Consequently, we discover additional plateau phases at $M/M_{\rm sat}$=1/3, 5/9, and 7/9, each distinctly different from the magnon crystal states.
\end{abstract}

\maketitle
\section{Introduction}
Quantum kagome antiferromagnets are renowned for hosting novel ground states due to the influence of frustration, garnering significant attention in the field of condensed matter physics~\cite{KL1,KL2,KL3}.
For the case of $S=1/2$, under zero magnetic field, various theoretical calculations predict the existence of quantum spin liquids with finite gaps between singlet and triplet states~\cite{KLZ2-1,KLZ2-2,KLZ2-3}, gapless quantum spin liquids~\cite{KLU1-1,KLU1-2,KLU1-3,KLU1-4,KLU1-5} and valence bond crystal (VBC) states~\cite{KLVBC1,KLVBC2,KLVBC3}. 
Moreover, under a magnetic field, theoretical investigations employing techniques such as exact diagonalization and the density matrix renormalization group methods have revealed the emergence of magnetization plateaus. These plateaus occur at $M/M_{\rm sat}$=1/9, 1/3, 5/9, and 7/9, where $M$ represents magnetization and $M_{\rm sat}$ denotes saturation magnetization~\cite{KLMH1,KLMH3,KLMH4,KLMH5,KLMH6,KLMH7,KLMH8,KLMH9}.
Among them, the 7/9 plateau is rigorously established to exist theoretically, corresponding to a state referred to as the magnon crystal state (MCS) or VBC state~\cite{MCS1}, where magnons are periodically localized on the lattice.
Similarly, it is anticipated that the 1/3 and 5/9 plateaus also correspond to the MCS~\cite{KLMH3,KLMH4,MCS2,CdCOHNOHO}.
The state of the 1/9 plateau remains theoretically contentious, predicted to be as either a $Z_3$ spin liquid~\cite{KLMH4} or a VBC~\cite{KLMH5,KLMH8} state, with detailed magnetic structure yet to be elucidated.
Very recently and interestingly, the 1/9 and 1/3 plateaus have also been experimentally observed in the kagome compound $ \rm YCu_3(OH)_{6+\it x}Br_{3-\it x}$ ($x\approx0.5$)~\cite{YCOHB1,YCOHB2,YCOHB3}.
Therefore, theoretical and experimental studies on magnetization plateaus in the kagome lattice are currently attracting significant attention.
These theoretical studies have primarily focused on the ideal kagome lattice where all exchange interactions are equivalent. However, in many kagome antiferromagnetic compounds, distorted kagome lattices are formed, resulting in non-equivalent exchange interactions~\cite{variKLMat}.
Nevertheless, theoretical investigations of distorted kagome lattices are scarce~\cite{DKL1,DKL2}. Therefore, it can be expected that new quantum states, which do not exist in undistorted kagome lattices, may emerge.

$\rm Y_3Cu_9(OH)_{19}Cl_8$ is a recently synthesized compound where copper ions ($\rm Cu^{2+}$) with $S=1/2$ spins form a distorted kagome lattice~\cite{YCOHC1,YCOHC2,YCOHC3,YCOHC4,YCOHC5,YCOHC6}.
There are three types of nearest-neighbor exchange interactions as shown in Fig.~\ref{F1}.
A previous study has theoretically revealed the existence of a ground state with ${\mathbf Q}=(1/3,1/3)$, coplanar long-range order in the classical spin system under zero magnetic field, which has also been experimentally observed by means of neutron scattering experiments. 
The candidate values for the exchange interactions have been proposed in two previous studies~\cite{YCOHC4,YCOHC6}. In this paper, we refer to them as Type~1~\cite{YCOHC4} and Type~2~\cite{YCOHC6}. However, the quantum magnetic properties of the spin system under a magnetic field have not been sufficiently investigated, and thus remain unclear. 
Therefore, it is important to investigate the ground state in the quantum spin model for $\rm Y_3Cu_9(OH)_{19}Cl_8$.

In this study, we investigate the ground state of this model under a magnetic field by exact diagonalization using the Lanczos method. We reveal that clear 1/3, 5/9, and 7/9 plateaus are present for both the Type~1 and Type~2 parameter sets. 
These ground states are identified as the MCS based on their spin structures.
Given the uncertainty in the exchange interactions of $\rm Y_3Cu_9(OH)_{19}Cl_8$, we perform additional  calculations by varying the exchange interactions within a certain range.
Consequently, we find additional plateau phases at $M/M_{\rm{sat}}=1/3$, 5/9, and 7/9, each distinctly different from the MCS.
We argue that clear magnetization plateaus can be observed through high-field magnetization measurements conducted on $\rm Y_3Cu_9(OH)_{19}Cl_8$.

\section{Model}
\label{sec2}
\begin{figure}[tb]
  \centering
  \includegraphics[width=80mm]{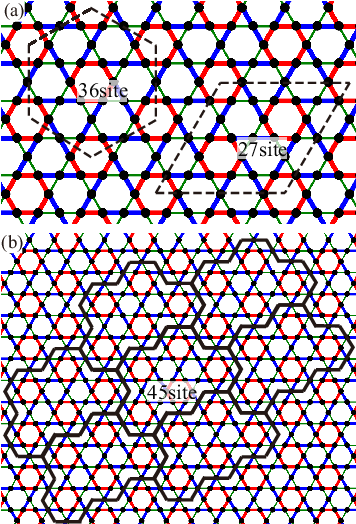}
  \caption{Lattice structure of the kagome lattice with ${\rm Y_3Cu_9(OH)_{19}Cl_8}$-type distortion. 
  The red solid, green thin, and  blue solid lines denote the exchange interactions $J_{\rm{r}},J_{\rm{g}}$, and $J_{\rm{b}}$, respectively.
  (a) The regular hexagon and rhombus represent clusters under PBCs with $N=36$ and $N=27$, respectively. 
  (b) The hexadecagon represents a cluster under PBCs with $N=45$. To clarify the concept of the PBCs, six hexadecagons are drawn around the central hexadecagon.
  \label{F1}}
\end{figure}
The Hamiltonian for the spin-1/2 kagome lattice with ${\rm Y_3Cu_9(OH)_{19}Cl_8}$-type distortion shown in Fig.~\ref{F1} in a magnetic field is defined as follows:
\begin{eqnarray} 
\mathcal{H} &=& \sum_{\langle i,j \rangle }J_{i,j}\mathbf{S}_i \cdot \mathbf{S}_j - g\mu_{\rm B}H\sum_i S^{z}_i,
\label{Hami}
\end{eqnarray}
where $\mathbf{S}_i$ is the spin-1/2 operator at the $i$-th site, $S^z_i$ is the $z$ component of $\mathbf{S}_i$, $\langle i,j \rangle$ runs over the nearest-neighbor  spin pairs of the kagome lattice, 
$J_{i,j}$ represents the nearest-neighbor exchange interactions, specifically, one of $J_{\rm r}$, $J_{\rm g}$, or $J_{\rm b}$ as shown in Fig.~\ref{F1},
$g$ is the gyromagnetic ratio,  $\mu_{\rm B}$ is the Bohr magneton, and $H$ represents the magnitude of the magnetic field applied in the $z$ direction.
${\rm Y_3Cu_9(OH)_{19}Cl_8}$ consists of three nearest-neighbor exchange interactions,  $J_{\rm r}$, $J_{\rm g}$, and $J_{\rm b}$, and the unit cell contains nine ${\rm Cu^{2+}}$ ions.
The values of the three exchange interactions are determined by the density functional theory (DFT) calculation~\cite{YCOHC4} and linear spin-wave analysis~\cite{YCOHC6}, referred to as the Type~1 and Type~2 in this paper, respectively.
The values of the exchange interactions for the Type~1 and Type~2 are listed in Table~\ref{Type12}.
\begin{table}[tb]
    \centering
    \caption{Candidate exchange interaction values of ${\rm Y_3Cu_9(OH)_{19}Cl_8}$.}
    \begin{tabular}{c c c c c} \hline 
              &\ \  & Type~1~\cite{YCOHC4}   &\ \  &Type~2~\cite{YCOHC6}  \\ \hline 
       $J_{\rm r}$  &\ \  & 134.2 K &\ \  &140 K \\ 
       $J_{\rm g}$  &\ \  & 8.7 K   &\ \  &63 K  \\ 
       $J_{\rm b}$  &\ \  & 154.4 K &\ \  &140 K \\
       $J_{\rm r}/J_{\rm b}$  &\ \  & 0.869 &\ \  & 1 \\ 
       $J_{\rm g}/J_{\rm b}$  &\ \  & 0.056   &\ \  & 0.45   \\          \hline 
    \end{tabular}
  \label{Type12}
\end{table}
We perform exact diagonalization using the Lanczos method for clusters with $N=27$, 36, and 45 under periodic boundary conditions (PBCs) as shown in Fig.~\ref{F1}, where $N$ represents the number of sites.
Our calculations for $N=45$ cannot be performed some subspaces with $M=1/2, 3/2, ..., 45/2$. For example, the $M=1/2$ subspace has the largest dimension of $\binom{45}{22}$, making the computation time and memory insufficient. 
Therefore, calculations are conducted within the range of $M \ge 23/2$ ($M/M_{\rm{sat}} \ge 23/45$), where the dimensions of these subspaces are $\binom{45}{11}$ or less.

In the following, we describe this model under four different limits. 
 When $J_{\rm r}=J_{\rm g}=J_{\rm b}$, it corresponds to the undistorted kagome lattice. 
 When $J_{\rm g}=J_{\rm b}=0$ and  $J_{\rm r}>0$, the model consists of isolated hexamers connected by $J_{\rm r}$ bonds and monomers. 
 When $J_{\rm r}=J_{\rm b}=0$ and  $J_{\rm g}>0$, the model consists of isolated trimers connected by $J_{\rm g}$ bonds. 
 When $J_{\rm g}=0$, $J_{\rm r}>0$ and  $J_{\rm b}>0$ the model exhibits frustration due to the presence of 9-site triangles, in which $J_{\rm r}$ bonds connect three trimers formed by $J_{\rm b}$ bonds. 
This means that both type~1 and type~2 models exhibit frustration.

\begin{figure}[tb]
  \centering
  \includegraphics[width=86mm]{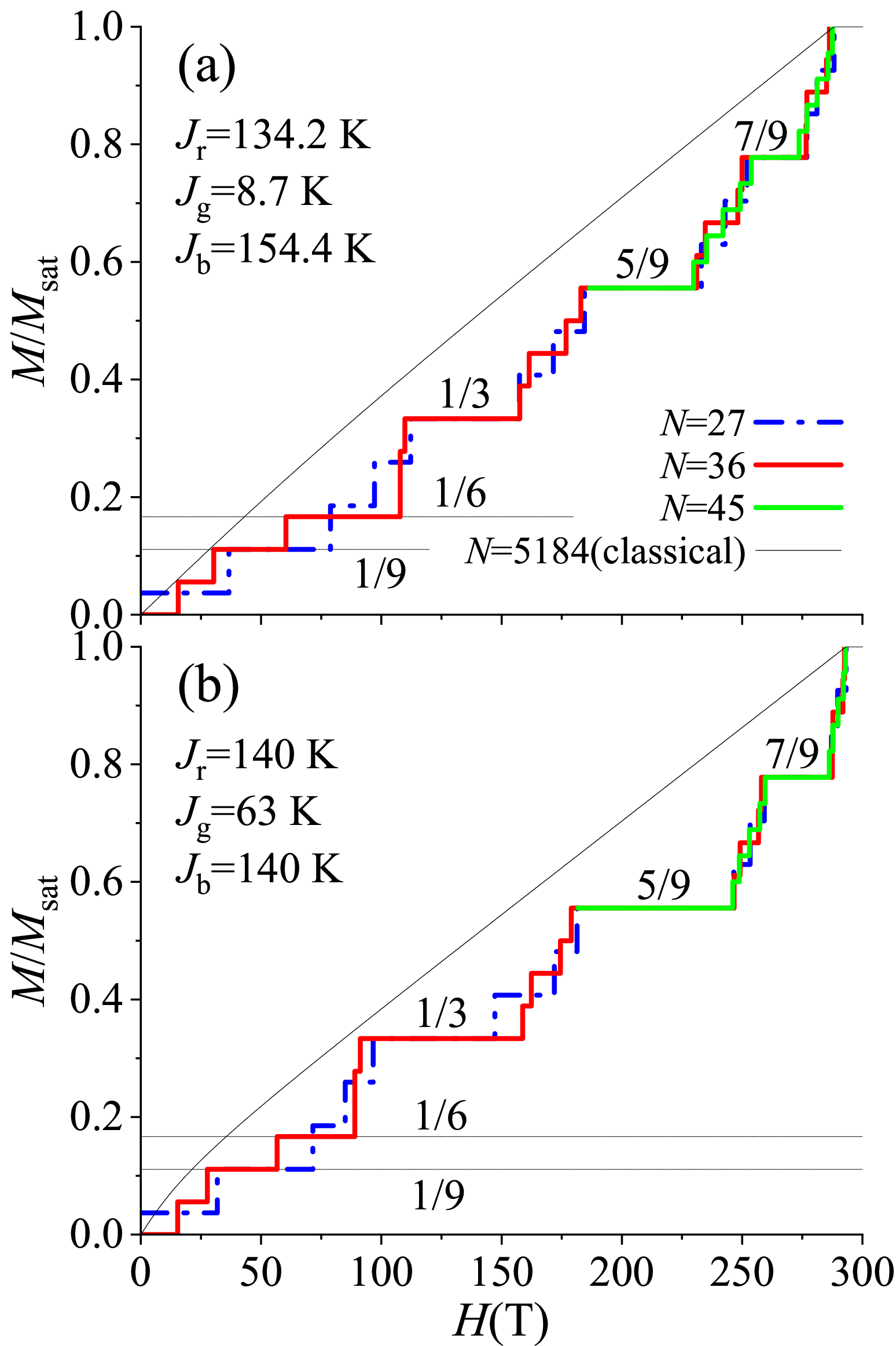}
  \caption{ Magnetization curves of the kagome lattice with (a) the Type~1 and (b) Type~2 parameters.
  The blue dashed, red solid, and green solid lines represent the calculation results for $N=27$, 36, and 45, respectively. The black solid lines show the calculation results for the classical spin system at $N=5184$. 
  Fractional values indicate the values of  $M/M_{\rm sat}$. The horizontal dashed lines are reference lines added for clarity.
  \label{F2}}
\end{figure}

\section{RESULTS and Discussions}
\label{sec3}
\subsection{Magnetization curve of $\rm{Y}_3\rm{Cu}_9(\rm{OH})_{19}\rm{Cl}_8$}
\label{subA}
Figure~\ref{F2}  depicts the magnetization curves at zero temperature of the Type~1 and Type~2 for ${\rm Y_3Cu_9(OH)_{19}Cl_8}$. Here, we set $g=2$. 
Both the Type~1 and Type~2 exhibit clear plateaus at $M/M_{\rm{sat}} =$ 1/3, 5/9, and 7/9.
In the Type~1, the widths of these plateaus show almost no dependence on the number of sites $N$, indicating negligible finite-size effects. Thus, these plateaus persist even in the thermodynamic limit.
In the Type~2, while the effects of size are negligible for the 5/9 and 7/9 plateaus, the width of the 1/3 plateau varies significantly at $N=27$ and $N=36$. However, the plateau width is broader for $N=36$ than for $N=27$, indicating its significant persistence even in the thermodynamic limit.
The magnetic structures of all three magnetization plateaus are confirmed to be the MCS in Sec.~\ref{subB}.
Figure~\ref{F2}  shows that all of these magnetization plateaus can be observed through magnetization measurements on $\rm{Y}_3\rm{Cu}_9(\rm{OH})_{19}\rm{Cl}_8$. Notably, the 1/3 plateau is observable at approximately 150~T, and the possibility of observing all plateaus increases at a stronger field of approximately 300~T, achievable with state-of-the-art magnetic field generators~\cite{600T1,600T2,600T3}.

The 1/9 plateau exhibits significant finite-size effects, with the plateau width being smaller for $N=36$ compared to $N=27$. Therefore, it is currently inconclusive whether the plateau width is finite in the thermodynamic limit based on our results. 
Further calculations with larger clusters, i.e., $N\geq45$, are needed to draw a conclusion about the finite-size effects of the 1/9 plateau.

In Fig.~\ref{F2}(a), the width of the 1/6 plateau at $N=36$ is comparable to that of the 1/3 plateau. 
Hence, the presence of the 1/6 plateau is also suggested. 
Since the $N=27$ system can only have values of $M/M_{\rm{sat}}=n/27$ ($n=1, 3, 5, ..., 27$), the 1/6 plateau cannot exist. However, the plateau at $M/M_{\rm{sat}}=5/27$, which is closest to 1/6, is not wide.
Therefore, there are insufficient results to conclude the existence of the plateau.
 The 1/6 plateau requires the number of sites to be a multiple of 12, and since the model has 9 sites per unit cell, $N$ must be a multiple of 36. Therefore, it is currently impossible to evaluate it through diagonalization.

Both the Type~1 and Type~2 show a magnetic jump immediately after $M/M_{\rm{sat}}=1/6$ at $N=36$. This jump occurs from $M/M_{\rm{sat}}$=3/18(1/6) to $M/M_{\rm{sat}}$=5/18, skipping $M/M_{\rm{sat}}$=4/18, suggesting the possibility of a first-order field-induced quantum phase transition. 
However, at $N=27$, no jumps, such as from $M/M_{\rm{sat}}$=5/27 to $M/M_{\rm{sat}}$=9/27 skipping $M/M_{\rm{sat}}$=7/27, are observed.
Thus, the jump observed at $N=36$ might be special due to finite-size effects. 
Further calculations with larger clusters are also necessary.

The solid black lines in Fig.~\ref{F2} represent the magnetization curve at $T=0$ for $N=5184$ ($24\times24\times9$) in the classical spin system, calculated using the simulated annealing technique with the Monte Carlo method. 
To verify the finite-size effects, we also calculated for  $N=729$ ($9\times9\times9$); however, no differences were observed in the results.
Therefore,  we can conclude that the results of the classical spin system presented in Fig.~\ref{F2} exhibit no finite-size effects.
In the classical system, no phase transitions exist except for the point reaching saturation magnetization. 
Therefore, it can be concluded that the magnetization plateaus and jumps obtained in this study are quantum phase transitions unique to quantum spin systems.

\begin{figure}[tb]
  \centering
  \includegraphics[width=60mm]{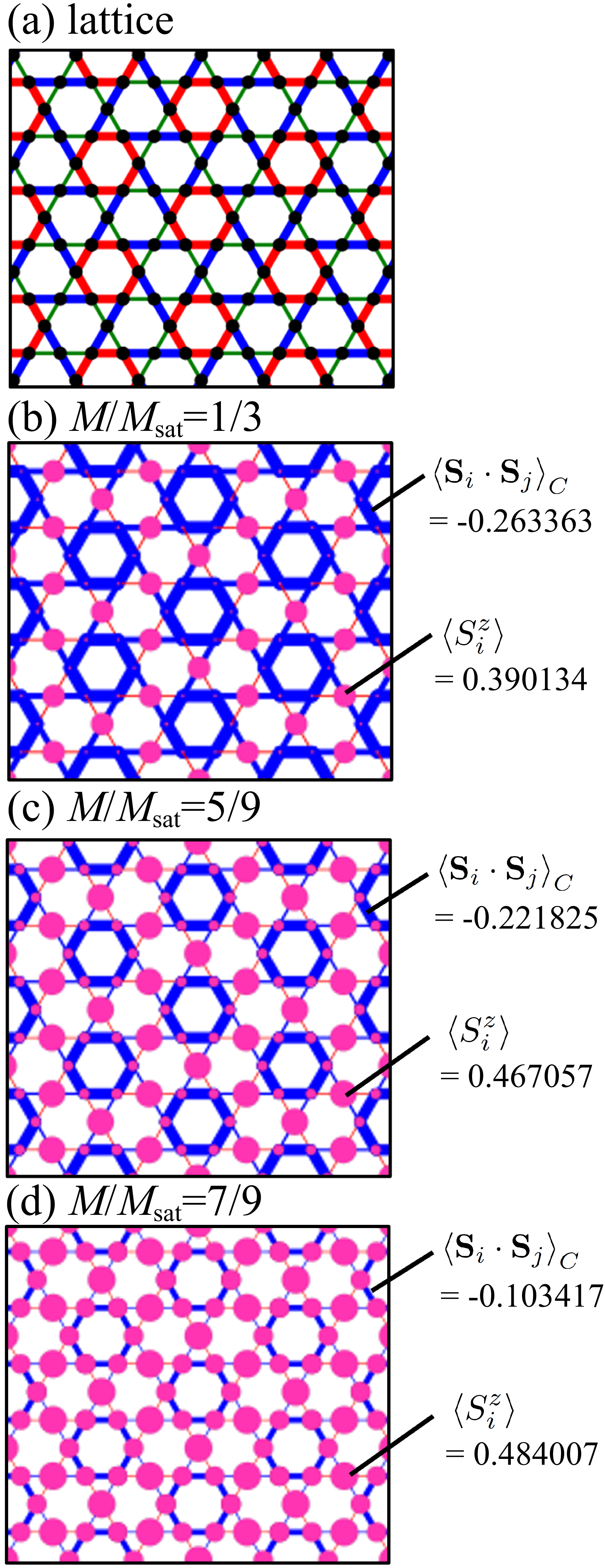}
  \caption{(a) Structure of the distorted kagome lattice. 
The red solid, green thin, and  blue solid lines denote the exchange interactions $J_{\rm{r}}, J_{\rm{g}}$, and $J_{\rm{b}}$, respectively.
Spin correlations of the kagome lattice with  $J_{\rm r}=140$~K, $J_{\rm g}=63$~K, and $J_{\rm b}=140$~K  at (b) $M/M_{\rm sat}$=1/3, (c) $M/M_{\rm sat}$=5/9, and (d) $M/M_{\rm sat}$=7/9. 
  (b),(c),(d) The blue and red lines represent negative and positive values of the spin-spin correlation $\Braket{\mathbf{S}_i \cdot \mathbf{S}_j}_C$, respectively, with the thickness of the lines corresponding to their magnitude. 
  The diameter of the magenta circles corresponds to the value of $\Braket{S_i^z}$ at that site.
  \label{F3}}
\end{figure}
\begin{figure*}[tb]
  \centering
  \includegraphics[width=176mm]{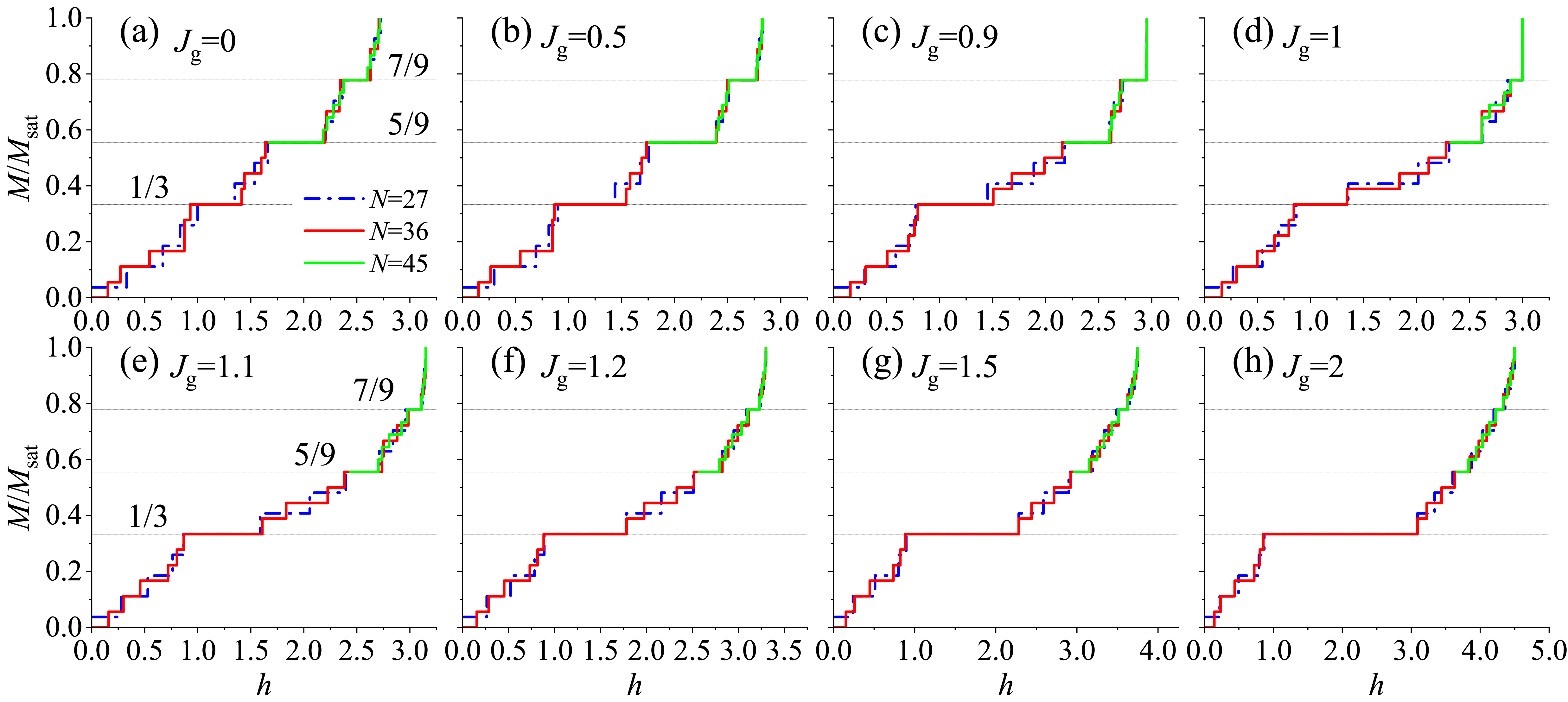}
  \caption{Magnetization curves for the kagome lattice with the exchange interactions $J_{\rm r}$, $J_{\rm g}$, and $J_{\rm b}$, where $J_{\rm r}$ and $J_{\rm b}$ are fixed at 1, and $J_{\rm g}$ varies from 0 to 2.
  The blue dashed, red solid, and green solid lines represent the calculation results for $N=27$, 36, and 45, respectively. 
  Fractional values indicate the values of  $M/M_{\rm sat}$. The horizontal dashed lines are reference lines added for clarity.
  \label{F4}}
\end{figure*}

\subsection{Spin structures of the magnetization plateaus}
\label{subB}
In this subsection, we investigate the magnetic structures of the 1/3, 5/9, and 7/9 magnetization plateaus.
Figure~\ref{F3} presents the calculation results for the spin structures of the 1/3, 5/9, and 7/9 plateaus for the Type~2 with $N=36$.
While we only show the calculation results for the Type~2 here, we have also performed the calculation for the Type~1, obtaining results similar to those of the Type~2.
Figure~\ref{F3}(a) illustrates the lattice structure, while Figs.~\ref{F3}(b), \ref{F3}(c), and \ref{F3}(d) depict the spin structures of the 1/3, 5/9, and 7/9 plateaus, respectively.
The blue and red lines in Figs.~\ref{F3}(b), \ref{F3}(c), and \ref{F3}(d) correspond to negative and positive of spin-spin correlations 
\begin{align*}
  \Braket{\mathbf{S}_i \cdot \mathbf{S}_j}_C = \Braket{\mathbf{S}_i \cdot \mathbf{S}_j} - \Braket{S_i^z}\Braket{S_j^z},
\end{align*}
 respectively, while the thickness of the lines corresponds to their magnitude.
The diameter of the magenta lattice points corresponds to the value of $\Braket{S_i^z}$ at that site.
The spin correlations between bonds connected by the exchange interaction $J_{\rm r}$ are strongest, forming periodically arranged hexagons. The remaining spins align parallel to the magnetic field. 
This structure corresponds to the MCS. 
Therefore, the magnetic structures of these plateaus are the same as the anticipated magnetic structures of the plateaus arising in the uniform kagome lattice~\cite{KLMH3,KLMH4,MCS1,MCS2,CdCOHNOHO}.
The emergence of the 1/3, 5/9, and 7/9 plateaus can be understood by considering a model with only the $J_{\rm r}$ hexagon. 
In this model, magnetization plateaus occur at $M=0,1,2$. 
These plateaus, combined with the remaining spins aligned parallel to the magnetic field, give rise to the $M/M_{\rm sat}$=1/3, 5/9, and 7/9 plateaus on the kagome lattice.
In the hexagon model, the exact values of the correlation $\Braket{\mathbf{S}_i \cdot \mathbf{S}_j}_C$ are -0.467 at $M=0$, -0.381 at $M=1$, and -0.194 at $M=2$. In the Type~2 model, the values of the correlation are -0.263 at $M/M_{\rm{sat}}=1/3$, -0.222 at $M/M_{\rm{sat}}=5/9$, and -0.103 at $M/M_{\rm{sat}}=7/9$, as shown in Fig.~\ref{F3}. These values in the Type~2 model are about 60\% of those in the hexagon model.
The presence of $J_b$ and $J_g$ weakens their correlations, but the correlation within the $J_r$ bonds remains significantly larger compared to those within the $J_b$ and $J_g$ bonds.

We also investigated  the spin structures of the 1/6 and 1/9 plateaus, but were unable to find any clues to the emergence of the plateaus.
Therefore, we refrain from further discussing the potential emergence of the 1/6 and 1/9 plateaus in this paper.

\begin{figure*}[tb]
  \centering
  \includegraphics[width=152mm]{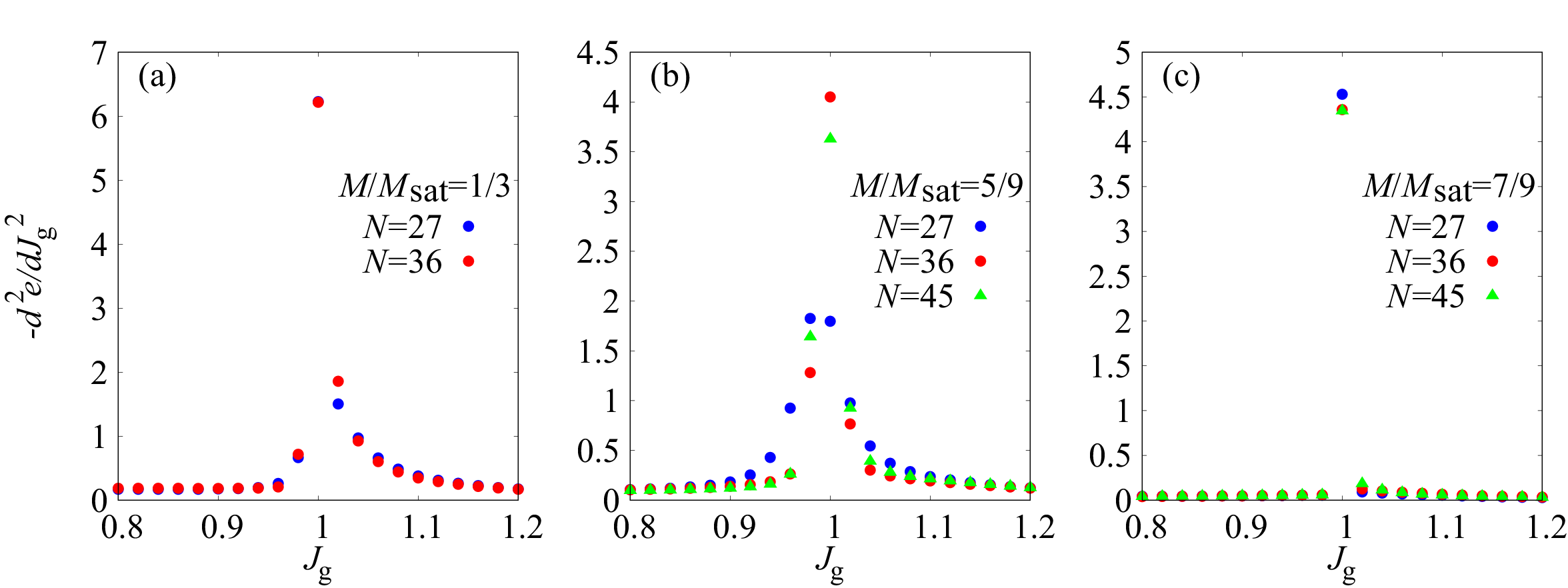}
  \caption{
Calculation results for the second derivative of the ground state energy per site, $e$, of the kagome lattice with $J_{\rm r}$ and $J_{\rm b}$ fixed at 1 and $J_{\rm g}$ as the variable parameter: (a) at  $M/M_{\rm sat}=1/3$, (b) at  $M/M_{\rm sat}=5/9$, and (c) at  $M/M_{\rm sat}=7/9$.
The blue circles represent the results for $N=27$, the red circles for $N=36$, and the light green triangles for $N=45$.
  \label{F5}}
\end{figure*}

\subsection{Calculations with other parameters}
\label{subC}
Two sets of exchange interaction parameters, the Type~1 and Type~2, listed in Table~\ref{Type12}, have been proposed for $\rm{Y}_3\rm{Cu}_9(\rm{OH})_{19}\rm{Cl}_8$.
 While these parameter sets are not significantly different, they do not completely match. Analyzing the model with wider range of parameters help us to get a deeper understanding of its magnetism.
Moreover, in the future, when new compounds are synthesized and their magnetic measurements are conducted, analyzing it based on alternative parameters would provide significant assistance.
This is important from a theoretical perspective as well, given the possibility of discovering new magnetic phases.
In the Type~1, $J_{\rm r} \approx J_{\rm b} \neq J_{\rm g}$ while in the Type~2, $J_{\rm r} = J_{\rm b} \neq J_{\rm g}$ as shown in Table~\ref{Type12}.
Therefore, in this section, we investigate the model with the restriction $J_{\rm r} = J_{\rm b} \neq J_{\rm g}$.
We consider the normalized Hamiltonian~(\ref{Hami}) with $J_{\rm r} = J_{\rm b}=1$ and the magnetic field term $h=g\mu_{\rm B}H$.

Figure~\ref{F4} shows the magnetization curves for $0 \leq  J_{\rm g} \leq 2$.
Interestingly, even at $J_{\rm g} =0$, the 1/3, 5/9, and 7/9 plateaus are observed, as evident from  Fig.~\ref{F4}(a). 
Furthermore, for $J_{\rm g} < 1$, these plateaus exhibit a broader width compared to that for $J_{\rm g} = 1$. 
This is because, in the region where $J_{\rm g}$ is small, $J_{\rm r}$ is relatively the largest, stabilizing the MCS, which is a state with strong correlations within the hexagons formed by $J_{\rm r}$ bonds.
Considering that the clear 5/9 and 7/9 magnetization plateaus have not yet been observed experimentally in the kagome lattice, the theoretical prediction of increased plateau widths provides sufficient motivation to conduct further experiments.
For $J_{\rm g} > 1$,  the 5/9 and 7/9 plateau widths narrow as $J_{\rm g}$ increases.
At $J_{\rm g}=1.5$, the plateau widths are significantly narrowed, as shown in Fig.~\ref{F4}(g), suggesting that these plateaus may not exist in the thermodynamic limit.
As $J_{\rm g}$ becomes larger, $J_{\rm r}$ becomes relatively smaller. Consequently, the MCS becomes unstable. Therefore, in the region where $J_{\rm g}$ is large, the 5/9 and 7/9 plateaus no longer exist.
In contrast, the width of the 1/3 plateau widens as $J_{\rm g}$ increases.
At $J_{\rm g}=2$, as depicted in Fig.~\ref{F4}(h), it is evident that the width of the 1/3 plateau significantly widens.
This is a natural consequence of the fact that when $J_{\rm g}$ becomes very large, it becomes equivalent to an isolated three-site model, resulting in the broadening of the 1/3 plateau.
Since in the region where $J_{\rm g}$ is large, the MCS becomes unstable,
it is expected that a phase transition exists around $J_{\rm g} =1$.
This will be discussed in detail later.

To confirm the presence or absence of phase transitions with respect to $J_{\rm g}$ at $M/M_{\rm sat}=1/3$, 5/9, and 7/9, we calculate the second derivative of the energy per site $e=E/N$ with respect to $J_{\rm g}$.
Figure~\ref{F5} shows the calculation results of  $-\frac{d^2e}{dJ_{\rm g}^2}$ at $M/M_{\rm sat}=1/3$, 5/9, and 7/9. 
It is evident that all three results exhibit sharp peaks at $J_{\rm g}=1$.
For $M/M_{\rm sat}=1/3$ and 5/9, based on the results in Figs.~\ref{F5}(a) and \ref{F5}(b), it is not possible to precisely determine the dimension of the phase transition, but it is expected to be either second-order or first-order. 
Further calculations with larger system sizes are required to clarify the dimension of the phase transition.
For $M/M_{\rm sat}=7/9$, as shown in Fig.\ref{F5}(c), it is apparent that $-\frac{d^2e}{dJ_{\rm g}^2}$ diverges at $J_{\rm g}=1$ sharply, indicating a first-order phase transition at $J_{\rm g}$.

\begin{figure}[tb]
  \centering
  \includegraphics[width=86mm]{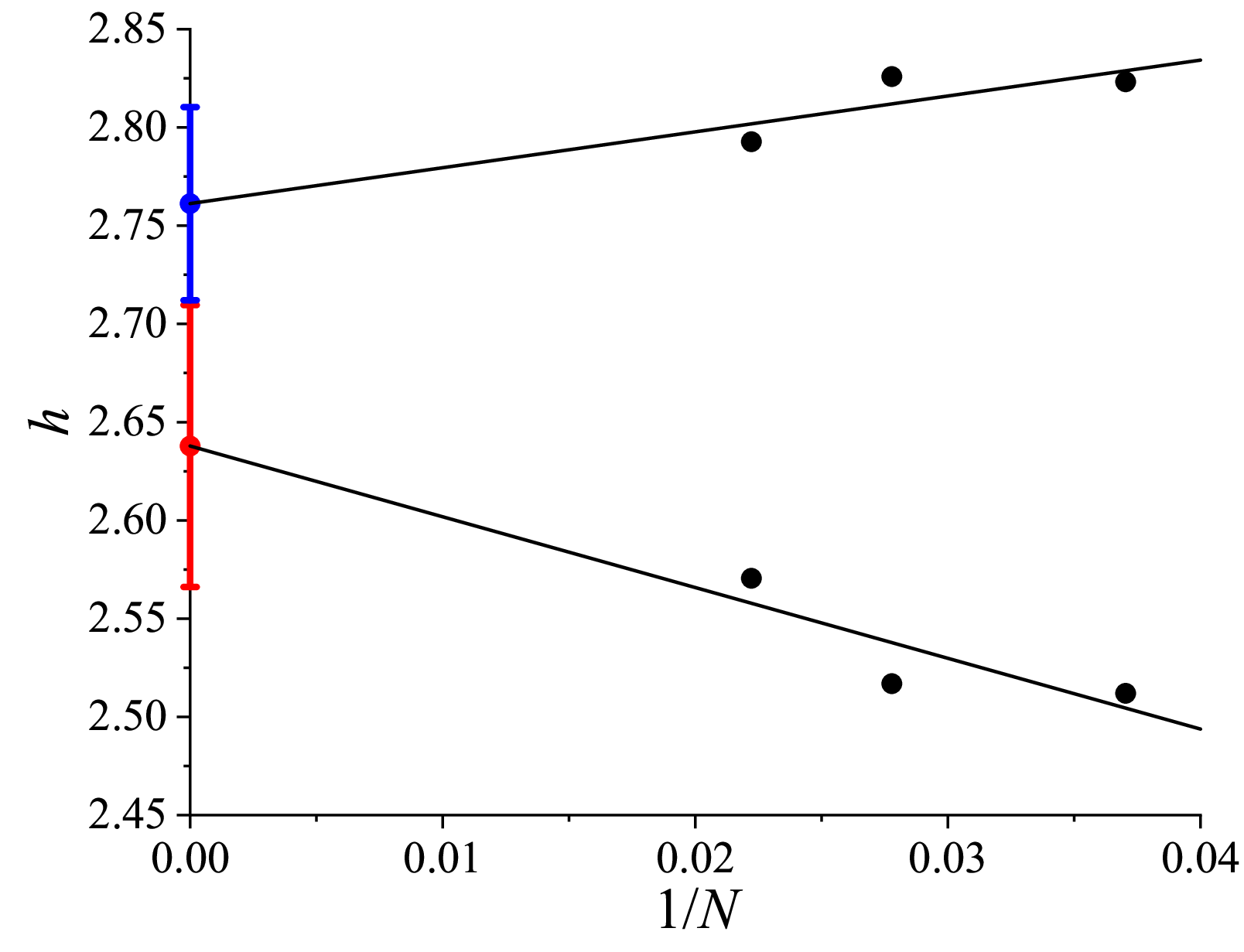}
  \caption{Results of finite-size scaling for the 5/9 plateau at $J_{\rm r} = J_{\rm b} =1$ and $J_{\rm g}=1.2$.
  The lower black line and upper black line represent the results of the linear fits for $h_{\rm l}$ and $h_{\rm u}$, respectively, which are defined as the lower and upper magnetic fields of the magnetization plateau.
  The red and blue points represent  $h_{\rm l}$ and $h_{\rm u}$ in the thermodynamic limit,  respectively, and  the bars represent the error from the linear fitting.
  \label{F6}}
\end{figure}

In the following we discuss the existence of plateaus in the thermodynamic limit.
The plateau width for $M/M_{\rm sat}=1/3$ remains broad regardless of $J_{\rm g}$, suggesting the existence of plateaus in the thermodynamic limit.
In contrast, for $M/M_{\rm sat}=5/9$ and $7/9$, particularly at $J_{\rm g} > 1$, the narrow plateau widths, as depicted in Fig.~\ref{F4}, make it difficult to discern the presence of plateaus in the thermodynamic limit.
Therefore, it is necessary to examine finite-size effects on the plateau width.
In such analyses, first, the existence of gapless excitations is assumed and an extrapolation is performed using the size dependence of $1/N$ or $1/N^{1/2}$~\cite{finiteSS1,finiteSS2}, where the former is used in this study. Let us define the lower magnetic field and upper magnetic field of the magnetization plateau are defined as $h_{\rm l}$ and $h_{\rm u}$, respectively.
 If $h_{\rm u} - h_{\rm l}$ remains positive in the thermodynamic limit $N \rightarrow \infty$, it is interpreted as an evidence of the presence of plateaus~\cite{finiteSS3,finiteSS4,finiteSS5}. 
 Figure~\ref{F6} shows the result of finite-size scaling for the 5/9 plateau at $J_{\rm g}=1.2$.
The black solid lines at the lower and upper sides correspond to $h_{\rm l}$ and $h_{\rm u}$, respectively.
The red and blue points represent  $h_{\rm l}$ and $h_{\rm u}$ in the thermodynamic limit, respectively, and error bars from the linear fitting are also shown. 
Here, the errors for $h_{\rm l}$ and $h_{\rm u}$ are denoted as $\delta h_{\rm l}$ and $\delta h_{\rm u}$, respectively. 
The estimates of finite-size effects from linear fitting are not completely accurate when the system size is not large. Therefore, in this paper, we cautiously  estimate that the plateau exists if the condition
\begin{align*}
  W_{\rm min} = \left(h_{\rm u} - \delta h_{\rm u}\right) - \left(h_{\rm l} + \delta h_{\rm l}\right) > 0
\end{align*}
 is satisfied. 
Table~\ref{Wmin1} and Table~\ref{Wmin2} present $W_{\rm min}$ for $M/M_{\rm sat}=5/9$ and 7/9, respectively, at various $J_{\rm g}$. 
For both $M/M_{\rm sat}=5/9$ and 7/9, $W_{\rm min}$ remains positive up to $J_{\rm g}=1.2$, indicating the presence of the plateaus for $J_{\rm g} \leq 1.2$.

\begin{table}[tb]
    \centering
    \caption{Results of the finite-size scaling for the plateau width at $M/M_{\rm sat}=5/9$.}
    \begin{tabular}{c c c c c c c} \hline 
       $J_g$  &\ \ & $h_{\rm l} \pm \delta h_{\rm l} $   &\ \ & $h_{\rm u} \pm \delta h_{\rm u}$  &\ \ & $W_{\rm min}$ \\ \hline 
       0.9  &\ \ & $2.18 \pm 0.06$   &\ \  & $2.61 \pm 0.03$ &\ \ & 0.331\\ 
       1.0  &\ \ & $2.32 \pm 0.09$   &\ \  & $2.62 \pm 0.01$ &\ \ & 0.202\\ 
       1.1  &\ \ & $2.47 \pm 0.09$   &\ \  & $2.71 \pm 0.07$ &\ \ & 0.079\\  
       1.2  &\ \ & $2.64 \pm 0.07$   &\ \  & $2.76 \pm 0.05$ &\ \ & 0.002\\  
       1.3  &\ \ & $2.78 \pm 0.06$   &\ \  & $2.85 \pm 0.03$ &\ \ & -0.020\\  
       1.4  &\ \ & $2.92 \pm 0.05$   &\ \  & $2.97 \pm 0.03$ &\ \ & -0.025\\ \hline 
    \end{tabular}
  \label{Wmin1}
\end{table}
\begin{table}[tb]
    \centering
    \caption{Results of the finite-size scaling for the plateau width at $M/M_{\rm sat}=7/9$.}
    \begin{tabular}{c c c c c c c} \hline 
       $J_g$  &\ \ & $h_{\rm l} \pm \delta h_{\rm l} $   &\ \ & $h_{\rm u} \pm \delta h_{\rm u}$  &\ \ & $W_{\rm min}$ \\ \hline 
       0.9  &\ \ & $2.70 \pm 0.04$   &\ \  & $2.95 \pm 0.001$ &\ \ & 0.207\\ 
       1.0  &\ \ & $2.93 \pm 0.03$   &\ \  & $3 \pm 0$ &\ \ & 0.038\\ 
       1.1  &\ \ & $3.01 \pm 0.04$   &\ \  & $3.09 \pm 0.02$ &\ \ & 0.025\\  
       1.2  &\ \ & $3.14 \pm 0.04$   &\ \  & $3.20 \pm 0.02$ &\ \ & 0.009\\  
       1.3  &\ \ & $3.27 \pm 0.03$   &\ \  & $3.33 \pm 0.02$ &\ \ & -0.001 \\  
       1.4  &\ \ & $3.41 \pm 0.03$   &\ \  & $3.46 \pm 0.02$ &\ \ & -0.007 \\ \hline 
    \end{tabular}
  \label{Wmin2}
\end{table}

\begin{figure}[tb]
  \centering
  \includegraphics[width=86mm]{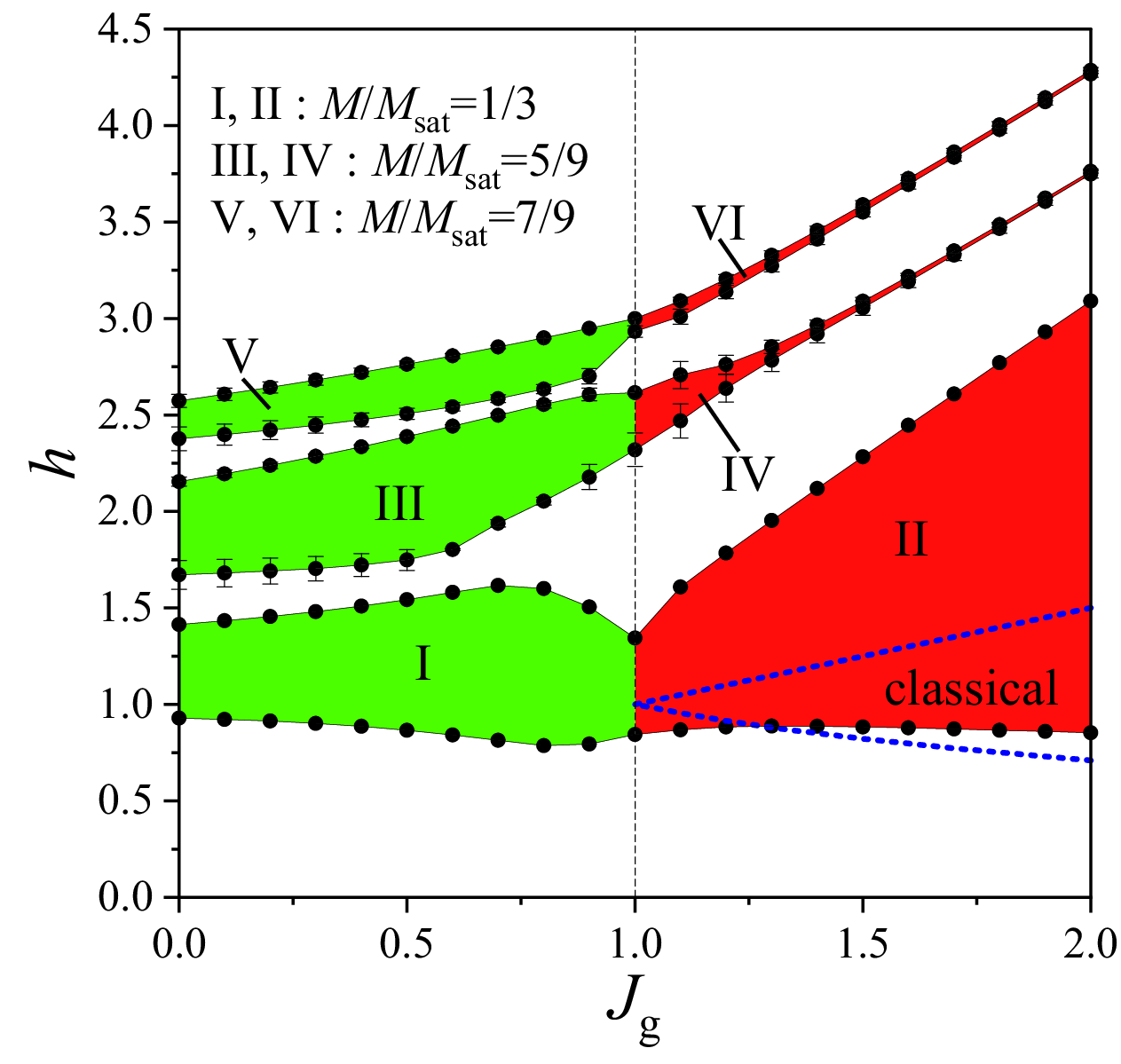}
  \caption{Magnetic phase diagram of the kagome lattice, focused on magnetization plateaus, with $J_{\rm r}=J_{\rm b}=1$.
There are a total of six plateau phases, each labeled with Roman numerals.
The plots with error bars for $M/M_{\rm sat}=5/9$ and 7/9 are estimates in the thermodynamic limit, obtained through the finite-size scaling for $N=27$, 36, and 45.
The plot for $M/M_{\rm sat}=1/3$ displays the results obtained for only $N=36$.
The area enclosed by the blue dashed lines represents the 1/3 plateau region in the classical spin system.
  \label{F7}}
\end{figure}

Since the widths of the plateaus and the phase boundaries have been clarified, we next discuss the phase diagram.
Figure~\ref{F7} depicts the phase diagram corresponding to each plateau, identifying a total of six plateau phases. 
As discussed above, the phase boundaries are at $J_{\rm g}=1.0$.
Consequently, there are the six plateau phases, labeled I to VI. 
For $M/M_{\rm sat}=5/9$ and 7/9, the finite-size scaling is conducted and error bars are included, whereas for $M/M_{\rm sat}=1/3$, results are shown for the $N=36$ system. 
This is because calculations were only performed for $N=27$ and 36, making it impossible to determine errors in finite-size scaling.
In all regions with respect to $J_{\rm g}$, there exists the broad 1/3 magnetization plateau.
Additionally, even when including errors from finite-size scaling, the 5/9 and 7/9 plateaus exist for $J_{\rm g}\leq1.2$.
For comparison, the region of the plateau phase in the classical spin system, calculated using the simulated annealing technique with the Monte Carlo method, is also depicted in Fig.~\ref{F7}. While the quantum system exhibits six distinct plateau phases, only one plateau phase appears in the classical system. Additionally, the region for the phase II in the quantum system is clearly broader than its counterpart in the classical system. This is due to quantum fluctuations that further stabilize the plateaus.
The phenomenon of plateau stabilization has been often observed in frustrated systems such as triangular lattices~\cite{TheTL1,TheAniTL1,TheAniTL2,TheAniTL3,ExpTL1}.

\begin{figure}[tb]
  \centering
  \includegraphics[width=86mm]{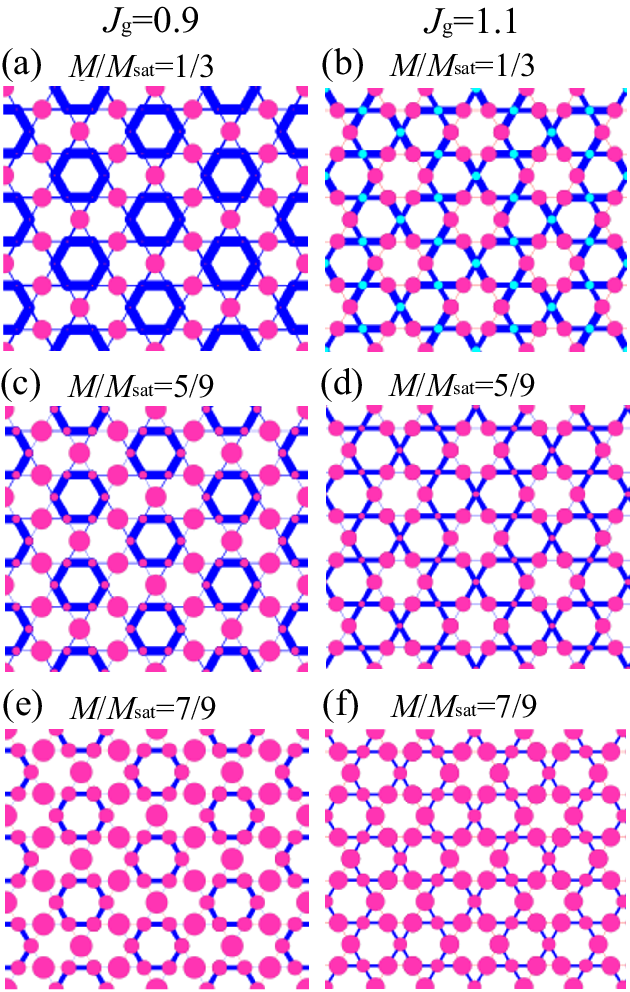}
  \caption{%
Spin correlations of each magnetization plateau phase in the kagome lattice at (a),(c),(e) $J_{\rm g}$=0.9, and (b),(d),(f) $J_{\rm g}$=1.1.
  The blue and red lines represent negative and positive values of the spin-spin correlation $\Braket{\mathbf{S}_i \cdot \mathbf{S}_j}_C$, respectively, with the thickness of the lines corresponding to their magnitude. 
Similarly, the cyan and magenta circles represent negative and positive values of the local magnetization $\Braket{S_i^z}$ at that site, respectively, with the diameter of the circles corresponding to their magnitude.
  \label{F8}}
\end{figure}

To investigate the magnetic structures of these six plateau phases, we calculate the spin correlations at  $J_{\rm g}=0.9$ and  $J_{\rm g}=1.1$. Figure~\ref{F8} presents the calculation results of the spin correlation functions.
 For $M/M_{\rm sat}=1/3$, 5/9, and 7/9, the spin structures at $J_{\rm g}=0.9$ and  $J_{\rm g}=1.1$ are distinctly different. 
Examining Figs.~\ref{F8}(a), \ref{F8}(c), and \ref{F8}(e), it is clear that at  $J_{\rm g}=0.9$, all three plateaus exhibit the MCS. 
As shown in Figs.~\ref{F8}(b), \ref{F8}(d), and \ref{F8}(f), the sites with the lowest value of $\Braket{S_i^z}$ at $J_{\rm g} = 0.9$ change significantly to exhibit the highest value at $J_{\rm g} =1.1$.
For $M/M_{\rm{sat}}=1/3$, Fig.~\ref{F8}(b) indicates that, solely focusing on $\Braket{S_i^z}$, the state is equivalent to the $\sqrt{3}\times\sqrt{3}$ up-up-down state, which is found to be stable for $J_{\rm g} > 1$ in our classical calculations.
However, the $M/M_{\rm{sat}}=5/9$ and 7/9 plateaus do not appear in the classical spin system.
Additionally, from the results shown in Fig.~\ref{F8}(d) and Fig.~\ref{F8}(f), it remains unclear why these plateaus stabilize in the quantum spin system.
In $J_1-J_2-J_3$ square kagome lattice model similar to the present model, a plateau phase with spontaneous rotational symmetry breaking has been found to exist in addition to the MCS (VBC) at $M/M_{\rm{sat}}=2/3$~\cite{SKL}.
It is possible that the present model stabilizes under similar principles; however, in the current calculations, no breaking of rotational symmetry has been observed. Further calculations with larger system sizes will be necessary, posing a challenge for future research.
\section{CONCLUSIONS}
\label{sec6}
The recently discovered spin-1/2 kagome antiferromagnet $\rm{Y}_3\rm{Cu}_9(\rm{OH})_{19}\rm{Cl}_8$ is predicted to exhibit the Type~1 and Type~2 exchange interaction parameters, as shown in Table~\ref{Type12}, according to separate studies.
Inspired by this discovery, we investigated the ground state of the spin-1/2 Heisenberg kagome model with three types of exchange interactions in the magnetic field using the Lanczos-type exact diagonalization method.
We confirmed the existence of clear $M/M_{\rm{sat}}=1/3$, 5/9, and 7/9 magnetization plateaus under both the Type~1 and Type~2 parameter sets, identified as the MCS. 
Our calculation results suggest that these plateaus could potentially be verified experimentally through magnetization measurements under the magnetic field of approximately 300~T, achievable with state-of-the-art magnetic field generators~\cite{600T1,600T2,600T3}.
Notably, clear 5/9 and 7/9 magnetization plateaus have not yet been observed in experiments, highlighting the importance of additional experimental investigation into $\rm{Y}_3\rm{Cu}_9(\rm{OH})_{19}\rm{Cl}_8$.
Additionally, by performing calculations of the kagome lattice with parameter sets other than the Type~1 and Type~2, we discovered additional plateau phases at $M/M_{\rm{sat}}=1/3$, 5/9, and 7/9.
The mechanism behind the 1/3 plateau can be easily explained by analytically solving a three-site system. In contrast, the mechanisms for the 5/9 and 7/9 plateaus could not be clarified in this study.
Further experimental and theoretical research is therefore needed to elucidate these phenomena.
 These findings indicate that further experimental studies on new kagome compounds may successfully observe these new plateaus.
\begin{acknowledgments}
We thank the Supercomputer Center, the Institute for Solid State Physics, the University of Tokyo for the use of the facilities.
\end{acknowledgments}
\bibliography{ref.bib} 
\end{document}